# Bandstructure Effects in Silicon Nanowire Electron Transport

Neophytos Neophytou, Abhijeet Paul, Mark Lundstrom and Gerhard Klimeck

School of Electrical and Computer Engineering, Purdue University, West Lafayette,

Indiana 47907-1285

Email: neophyto@purdue.edu

## ABSTRACT

Bandstructure effects in the electronic transport of strongly quantized silicon nanowire field-effect-transistors (FET) in various transport orientations are examined. A 10-band $sp^3d^5s^*$ semi-empirical atomistic tight-binding model coupled to a self consistent Poisson solver is used for the dispersion calculation. A semi-classical, ballistic FET model is used to evaluate the current-voltage characteristics. It is found that the total gate capacitance is degraded from the oxide capacitance value by 30% for wires in all the considered transport orientations ([100], [110], [111]). Different wire directions primarily influence the carrier velocities, which mainly determine the relative performance differences, while the total charge difference is weakly affected. The velocities depend on the effective mass and degeneracy of the dispersions. The [110] and secondly the [100] oriented 3nm thick nanowires examined, indicate the best ON-current performance compared to [111] wires. The dispersion features are strong functions of quantization. Effects such as valley splitting can lift the degeneracies especially for wires with cross section sides below 3nm. The effective masses also change significantly with quantization, and change differently for different transport orientations. For the cases of [100] and [111] wires the masses increase with quantization, however, in the [110] case, the mass decreases. The mass variations can be explained from the non-parabolicities and anisotropies that reside in the first Brillouin zone of silicon.

**Index terms – nanowire, bandstructure, tight binding, transistors, MOSFETs, non-parabolicity, effective mass, injection velocity, quantum capacitance, anisotropy.**

# 1. Introduction

As transistor sizes shrink down to the nanoscale, CMOS development investigates alternative structures and devices [1]. Existing CMOS field-effect transistors are expected to evolve from planar, to 3D non-planar devices at nanometer sizes. A possible device approach that has attracted large attention recently because of its possibility of enhanced electrostatic control, is the use of nanowire (NW) transistors as field effect devices. Nanowire transistors of diameters even down to 3nm have already been demonstrated by various experimental groups [2-4]. Under extreme scaling of the device's dimensions, the atoms in the cross section will be countable, and crystal symmetry, bond orientation and quantum mechanical confinement will matter. Proper atomistic modeling is therefore essential in understanding the electrical characteristics of ultra-scaled cross section nanowire devices. This work identifies the main bandstructure parameters that influence the transport properties of nanowire devices by using a nearest-neighbor tight binding (TB) model ($sp^3d^5s^*$) [5-8] for electronic structure calculation, coupled to a 2D Poisson solver for electrostatics. To evaluate transport characteristics, a simple semi-classical ballistic model [9, 10] (Fig. 1a) is used. The ballistic transport characteristics of square nanowires of 3nm width, oriented in [100], [110] and [111] transport directions are examined and compared.

The electrostatic potential variations in the lattice are calculated using a 2D Poisson solver. Self-consistently accounting for the potential variations in the nanowire cross section is a critical step in evaluating nanowire transport characteristics. Although this factor causes only small shifts of the energy levels, and weak lifting of degeneracies, it strongly influences the charge placement in the cross section of the wire. The charge placement, together with the small 1D density of states in nanowires, strongly degrades the capacitance of the device by up to 30% and affects its performance. This degradation is the *same* for all wire orientations. It is shown that the carrier injection velocities (controlled by the transport mass) and the degeneracies (controlled by orientation) are the dominant determining factors of the relative device performance for different wire orientations. In terms of ON-current capabilities, transport at small nanowire dimensions will be preferable in the [110] oriented devices that have the lowest mass and highest carrier velocities, closely followed by the [100] devices with a little higher masses. [111] nanowires indicate lower performance due to their much heavier transport masses in agreement with [11]. As also shown by other authors, the masses and degeneracies are strong functions of not only directionality, but also of structural quantization [12-14]. Specifically, for the case of mass variations, this work shows how quantization affects the masses differently in different transport orientations. It is a result of the non-parabolicity and anisotropy of the Si bandstructure that is particularly evident in strongly quantized nanowires. Wires in [110] transport orientation experience a reduction in the electron transport effective mass compared to [100] or [111] oriented wires. A simple analytical approach presented in this work, provides insight in understanding these variations. Other effects, such as valley splitting, also have directional dependence and are significantly enhanced in [110] nanowires compared to [100] or [111] nanowires.

*Necessity of atomistic modeling:* The effects investigated in this study, are mainly atomistic effects, which the usual effective mass approximation (EMA) fails to capture. When comparing devices in different orientations, atomistic description of the device has the advantage of being able to automatically capture the valley projections and extract the dispersions of the nanowires in the transport orientation. Atomistic modeling also automatically includes information about band coupling and mass variations as functions of quantization. The problem of identifying the correct bandstructure and effective masses of nanowires has been addressed by various authors in references [15-18] with qualitative agreement on the main features of the electronic structures. Other sophisticated techniques for electronic structure calculation also mention mass variations in nanostructures from their bulk values, which results in different threshold voltages and ON-current densities [15, 19]. By adjusting the effective masses to map masses extracted from atomistic calculations, however, especially for [100] oriented nanowires, and at some cases for other orientations, the EMA can still be used [15, 20] for the conduction band only. In general, however, this method is not always valid, and atomistic simulations are more appropriate for nanowires of a few nanometers of cross sectional sides.

*This paper is organized as follows:* In section 2, the TB model, its validity and the simulation approach is described. Section 3, contains the numerical results. Section 3(a) examines the behavior of bandstructure under charge filling of the lattice for nanowires in different orientations for 3nm cross section square nanowires. Section 3(b) compares the performance of the nanowires in terms of total gate capacitance, quantum capacitance, injection velocity and drive current capabilities for nanowires in different orientations. Explanations for the relative performance are given in terms of the most important dispersion properties (masses and degeneracies). Section 3(c) examines how structural quantization will impact the dispersions of wires with different cross sectional areas. The valley splitting and the mass variation in wires of different cross sections are examined. Finally, in 3(d) an explanation of the dispersion mass and band edge variation is given using extracted bands from the bulk bandstructure. Section 4 summarizes and concludes the paper.

## 2. Approach

*Motivation for an empirical TB model:* At the nanometer scale the concept of a "new device" and a "new material" are blurred. Quantum mechanics of the electronic structure, crystal symmetry, atomic composition and spatial disorder are important. A certain electronic structure model needs to satisfy several requirements to accurately capture nanoscale device physics. The finite extend of the devices rather than the infinite periodic nature speaks for the choice of a local basis set rather than a plane wave basis set. The stability of the bands in typical semiconductor devices speaks for a reduced model that takes the existence of bands for granted. The need to model complicated man-made heterostructures speaks for a nearest neighbor model to eliminate ambiguities of long-range coupling elements. The need to simulate large extensive structures containing tenths of millions of atoms [7], requires a reduced order model. The need to accurately model bandgaps (within a few meV) and masses (within a few %) speaks for an empirical bandstructure model rather than an ab-initio model. All these requirements have led to the choice of empirical TB in this work.

*The sp³d⁵s\* TB model:* The basis set of the sp$^3$d$^5$s\* nearest neighbor TB model used in this work, is composed of orthogonal localized orbitals. This type of basis makes it very attractive for accurate electronic structure of truncated nanostructures of finite sizes and composition variations on the nm-scale. It is a very convenient method to treat material and potential variations as well as strain fields at the nanoscale. The parameterization was performed using a genetic algorithm in [5], and the parameters extracted can reproduce the band edges of the bulk silicon bandstructure over the entire Brillouin zone. The model is described in detail in references [5-8]. The energy bands obtained for nanowires, as well as in the bulk case in energy regions away from the bulk minima, are in good quantitative agreement with other theoretical calculations using pseudopotential and ab-initio GW methods [21]. In this work, the electrostatic potential for charge self consistency is also included on the on-sides of the Hamiltonian in an effective potential approach which shifts the bands with no further change in connectivity.

*Validation of the model through experimental data:* Since the accuracy of the results presented here strongly depends on the validity of the TB model used, and especially on its transferability to nanostructures, it is convincing to mention that the same model and calibration parameters were used to explain experimental data in a variety of applications with excellent qualitative agreement. Some examples include explaining resonant tunneling diode applications for transport under high bias with charge self consistency [22-24], explaining experimental data for the bandgap of ultra scaled nanowires [25-27], valley splitting of tilted and disordered quantum wells [28], and the electronic structure of silicon systems with phosphorus impurities [29]. Further theoretical work presented in reference [30] examines the performance of core shell nanowires and validates against experimental data. Specifically, the theoretical calculation of experimental measurements of the bandgap of ultra scaled [112] oriented nanowires in [25-27] is a strong validation that the model captures the essential non-parabolicities in a large part of the Brillouin zone of Si. As it will be shown later on, non-parabolicites and anisotropies at high energies strongly influence the masses and band edges of nanowires. Since the bandgap of quantized systems is a strong function of the quantization masses in the two transverse directions, a verification of the experimentally deduced nanowires' bandgap, supports the theoretical prediction for the behavior of the wires' masses under strong quantization, and in extent the validity of the model.

*The simulation approach:* The model device simulated is a rectangular nanowire of 3nm x 3nm dimensions in various transport orientations. Unless otherwise stated, the specified wire orientation is the transport direction. Three different orientations, ([100], [110] and [111]) are investigated. The atomic arrangement in each case is different as shown in Fig. 1b.

The simulation procedure consists of three steps as shown in Fig. 1a and described below:

1. First, the bandstructure of the wire is calculated using an atomistic tight-binding model. In this case, each atomic side in the zincblende lattice is represented by a $sp^3d^5s^*$ basis in the wire Hamiltonian. Since only the conduction band is treated in this paper, the spin-orbit coupling is ignored. This approximation favors computational efficiency, without affecting the accuracy of the results [16]. The atoms that reside on the surface of the nanowire are passivated in the $sp^3$ hybridization scheme [31]. This technique successfully removes all dangling bonds which otherwise will create surface states with eigen-energies in the bandgap of the device. Any effect of surface reconstruction or surface imperfections is not considered in this study. Only the channel atoms enter the atomistic calculation in the Hamiltonian construction. At this step, the energy of the dispersion states and their wavefunctions are computed. Bandstructure effects such as valley splitting and effective mass change under physical quantization are investigated at this step for the nanowire of interest, using the equilibrium dispersion (flat electrostatic potential in the Hamiltonian).

2. A semiclassical top-of-the-barrier ballistic model is used to fill the dispersion states and compute the transport characteristics [9, 10]. This model assumes that the positive going states are filled according to the source Fermi level, whereas the negative going states according to the drain Fermi level. Once the occupancy of the dispersion states is computed, using their wavefunction from step 1, the charge distribution in each of the orbital sites of the system (and therefore the spatial distribution of charge) is obtained.

3. Using the charge distribution obtained in step 2, the 2D Poisson equation is solved in the cross section of the wire to obtain the electrostatic potential. The Poisson's equation is solved in 2D and all the atomic locations are collapsed on the 2D plane [32]. The Poisson domain is described by a finite difference mesh and contains the nanowire core on an atomistic mesh, the dielectric and the metal. The electrostatic potential is added to the diagonal elements of the atomistic Hamiltonian for recalculating the bandstructure until self consistency is achieved. In this step, the oxide is in all calculations is assumed to be $SiO_2$ of 1.1nm thickness. This dielectric is not included in the Hamiltonian, but only treated in the Poisson equation as a continuum medium. Any effects due to the potential variations along the transport direction are ignored. This falls under the assumption that at the ballistic limit the carrier injection at the top of the barrier is of most importance to the transport properties of the device.

Although the transport model used is simplistic, it allows for examining how the bandstructure of the nanowire alone will affect its ballistic transport characteristics, ignoring any short channel effects or quantum mechanical tunneling under the potential barrier. The same conclusion to this work can be obtained from full 3D quantum (NEGF) simulations [11], but the simple model used here provides physical insight. It is also mentioned that the main conclusions of this work will be valid for other nanowire cross sections, i.e. cylindrical, since the electronic properties of nanowires are a much more sensitive function of the quantization size rather than the quantization shape [15].

## 3. Results and Discussion

*(a) Effect of potential variations on the NW dispersion and charge distribution*

*Description of the dispersion in [100] oriented wires:* The dispersion of a [100] oriented nanowire is shown in Fig. 2c. It has a four-fold degenerate valley at the Γ point ($k_x$=0) resulting from the k-space projection of the four silicon ellipsoids that reside in the plane of quantization (here the *y-z* plane). There are two more valleys residing off-Γ (one in the positive and one in the negative $k_x$ axis), that result from the two off-plane ellipsoids. The first four appear lower in energy because of their heavy quantization mass ($m_y \sim m_l$=0.89$m_0$ and $m_z \sim m_t$=0.19$m_0$) and have lighter transport mass ($m_x \sim m_t$=0.19$m_0$). The other two appear at higher energies because of the lighter quantization masses ($m_y \sim m_t$=0.19$m_0$ and $m_z \sim m_t$=0.19$m_0$) and have heavier transport mass ($m_x \sim m_l$=0.89$m_0$). (The wire masses $m_x$, $m_y$, $m_z$ are close, but not exactly the bulk longitudinal and transverse masses for reasons that will be addressed later on).

*Change of the [100] wire dispersion due to potential variations / charge filling:* The first part of the results section investigates how potential variations in the cross section of a wire can change the dispersion and how the wavefunction shape changes as the lattice fills up with charge. Figure 2 shows device features for a 3nm square [100] oriented nanowire under low and high gate biases. (The drain bias used is $V_D$=0.5V in all cases throughout this work). Under low gate biases, the lattice is almost empty of charge (Fig. 2a) and the dispersion relation (Fig. 2c) is the equilibrium dispersion. Under high biases, there is significant charge filling of the lattice as shown in Fig. 2b. The charge distribution takes the shape of the underlying atomic positions. In these simulations, even under high inversion conditions, the wavefunction is pushed almost 0.5nm away from the Si/SiO$_2$ interface. The dispersion of this small size nanowires, on the other hand, is usually considered to be a material parameter, and under strong confinement a property of the geometry, but independent of charge filling of the lattice. It is shown, however, in Fig. 2d, that charge filling of the lattice causes changes in the dispersion of the nanowire even at the 3nm wire length scale. Here, the excited states at Γ shift down, and reside now below the off-Γ point valleys. In this case the change in the dispersion is small, but since it is associated with the wavefunction shape that gives rise to the charge distribution in the wire cross section, it can affect the devices capacitance and to some extent its transport characteristics.

*Change of the [110] wire dispersion due to potential variations / charge filling:* The change in the dispersion under potential variations is also observed in different wire orientations, which have different dispersion relations. The position of the bands shifts and degeneracies can also be lifted. Figure 3a-b shows the *E(k)* of a [110] oriented nanowire under low and high biases. The dispersion looks different from the [100] dispersion, with a two-fold degenerate band at Γ, and pair of two-fold degenerate bands off-Γ. A larger variation in the dispersion under charge filling of the lattice is observed compared to the [100] wire case. The band degeneracies are lifted (from 2 to 1) by several *meV*. This is an effect that cannot be captured in a simple EMA treatment.

*Change of the [111] wire dispersion due to potential variations / charge filling:*

Figures 3c-d show the same features for a [111] oriented wire. The degeneracy of the bands of this wire is 3 (for each valley) because of the symmetry between the transport axis (or equivalently the quantization plane in the perpendicular direction) and the three pairs of ellipsoids in the Si bandstructure. High biases increase band coupling, which slightly lifts the degeneracies. It is noted that in the case of [100] and [110] wires the conduction band minima is located at the Γ point since the quantized Δ valleys project there. In the [111] case, however, the conduction band minimum is located at 0.37 of the Brillouin zone (normalized to 1) as seen in Fig.3c,d for reasons explained in [33].

*Charge / velocity are invariant to self-consistency:* Just by looking at these variations in the dispersion, however, it is not clear that these will result in changes in the transport characteristics. Indeed, Fig. 3e-f compares the density of states and velocities at the same $E_f$-$E_c$ (difference of the Fermi level from the conduction band edge) between the equilibrium dispersion and the dispersion at various biases and little difference is observed. Quantities for two cases are calculated: (a) The Fermi level "scans" the equilibrium bandstructure and the charge and injection velocities are extracted, and (b) the results are extracted from the self-consistent calculations with potential variations in the lattice taken into consideration. The charge and injection velocity is plotted as a function of $E_f$. ($E_c$ is shifted to zero for all wires). There is no significant difference in these extracted quantities due to the potential variations, and the self-consistent vs. non-self-consistent curves fall almost on top of each other. For this example a large drain bias ($V_D$=0.5V) is used. Under low drain biases ($V_D$=1meV) and low temperatures, however, where the transport energy window can be comparable or even smaller than the changes in the bandstructure, evidence of the bandstructure differences in these two quantities as well as other quantities such as the transconductance are more likely to appear.

*Charge distribution is strongly dependent on self-consistency:* Although the charge and velocity appear to be only weakly modified by the self-consistent calculation, the self consistently extracted bandstructure corresponds to a different wavefunction shape which reflects to a different charge distribution in space. This is the quantity that causes degradation of the total gate capacitance as will be shown later and affects the transport characteristics, and not the dispersion changes by themselves. One therefore, has to also consider the change in the wavefunction that is associated with the dispersion changes. (In an earlier work, [34], it is shown that the current-voltage characteristics can be significantly overestimated if the spatial variation of the charge is not considered).

*Orientation differences in the charge:* The fact that the charge in Fig. 3e for any position of the Fermi level is always the highest in the [111] wire case, is due to the higher density of states and valley degeneracy. This particular wire orientation has the valleys with the heaviest mass ($0.47m_0$, where $m_0$ is the free electron mass) and the largest degeneracy ($D=6$). Therefore, at a certain energy level ($E_f$-$E_c$), there are more states occupied compared to the other wires. The [100] wire with mass $0.27m_0$ and $D=4$ of the lowest valleys, follows. The [110] wire has the lowest charge density at a certain energy level because of its lighter mass ($0.16m_0$) and lower degeneracy ($D=2$) at $\Gamma$.

*Orientation differences in the velocity:* The reverse trend is observed in Fig. 3f, where [110] wire has the highest velocity due to its lighter mass ($0.16m_0$). As higher k-states are occupied, the velocity increases since it is proportional to the slope of the bands. Noticeable here, is the fact that the carrier velocity in the [100] wire approaches the [110] velocity as the Fermi level is pushed into the conduction band. The lighter masses ($0.16m_0$) of the two-fold $\Gamma$ valleys in the [110] wire give an initial advantage over the heavier ($0.27m_0$) [100] wire $\Gamma$ valley masses. Once the heavier four-fold degenerate off-$\Gamma$ valleys (with mass $0.61m_0$) of the [110], and the heavy two-fold degenerate off-$\Gamma$ valleys (with mass $0.94m_0$) of the [100] start to populate, the carrier velocities become comparable in the two cases. The exact reasons why the masses have these values will be addressed later on in the paper, however this analysis can guide through the reasons why wires in different transport orientations have different properties.

*(b) Device performance comparison of NWs in different orientations*

One of the points made in the previous paragraph, are comparisons of the different wire orientations at the same Fermi level position into the dispersion of the wires. Although this is a rough estimate of the wires' properties, the Fermi level is not at the same position for all devices, except under special cases. In this section, the full self consistent model is implemented to compare the performance of the nanowires. Figure 4 shows a performance comparison between the wires in the [100], [110] and [111] orientations. The various performance quantities shown further on, are all compared at the same OFF current ($I_{OFF}$) for all devices.

*Gating induces same capacitance / charge in all wire directions:* Figure 4a shows the total gate capacitance ($C_G$) vs. gate bias ($V_G$) of the three wires at the same $I_{OFF}$. The total capacitance in the three wires is very similar for all gate biases for reasons we will explain later on. However, this is an indication that the same amount of inversion charge is accumulated in all wires irrespective of their orientations. Our calculation supports this argument too, showing that the charge difference between the wires at high inversion does not exceed 2%. In a relative performance comparison for wires in different orientations, therefore, the amount of charge will not affect the relative performance.

*Low semiconductor capacitance ($C_S$) degrades the gate capacitance ($C_G$) by 30%:* It is important to notice that for all three wire cases, the capacitance value is degraded from the oxide capacitance by almost 30%. This is an amount that corresponds to an effective increase in the oxide thickness of 0.54nm, which is 50% of the physical gate oxide thickness ($t_{ox}$=1.1nm). This large gate control reduction is evidence of low semiconductor capacitance ($C_S$) in low dimensional channels. The gate capacitance of a device is the series combination of the oxide capacitance ($C_{OX}$) and the semiconductor capacitance ($C_S$) given by the simple relation $C_G = \dfrac{C_S C_{OX}}{C_S + C_{OX}}$. For an electrostatically well behaved MOSFET device, $C_S$ should be an order of magnitude larger than $C_{OX}$ so that the $C_G$ and therefore the charge in the device is totally controlled by the gate. In this example, the oxide capacitance of the rectangular structure is 0.483 nF/m, numerically calculated using a 2D Poisson solver that takes the fringing at the edges into consideration. With $C_G$=0.3nF/m (maximum value of Fig. 4a), $C_S$ can therefore be computed to be $C_S = 0.8 nF/m$, which is only twice the value of the oxide capacitance (less than an order of magnitude difference).

*Cs controlling factors: Charge distribution peak, small $C_Q$:* $C_S$ is defined as the differential of the charge in the device with respect to the surface potential ($\psi_S$). In 1D systems, under a single band effective mass approximation, the charge is the integral of the 1D density of states ($g_{1D}$) convoluted with the Fermi function ($f(E_f-E)$) over energy as:

$$C_s = \frac{\partial(qn_s)}{\partial \psi_s} = \frac{\partial}{\partial \psi_s}\left(\int qg_{1D}f[(E_f - E_c - \varepsilon_i)/k_BT]dE\right) \qquad \text{Eqn. 1a}$$

where $q$ is the charge of the electron, $\psi_s$ is the surface potential, $E_f$ is the Fermi level, $E_c$ is the conduction band edge and $\varepsilon_i$ is the distance of the $i^{th}$ quantized subband above $E_c$ in energy. Carrying on the integration, the equation above results in:

$$Cs = \frac{q^2}{\pi}\left(\frac{2m}{\hbar^2}\right)^{1/2}\sqrt{k_BT}\sqrt{\pi}\frac{\partial}{\partial q\psi_s}\left(\mathfrak{I}_{-1/2}[(E_f - E_c - \varepsilon_i)/k_BT]\right) \qquad \text{Eqn. 1b}$$

$$= q^2\left(\frac{2m}{\pi\hbar^2}\right)^{1/2}\sqrt{k_BT}\,\mathfrak{I}_{-3/2}[(E_f - E_c - \varepsilon_i)/k_BT]\left(1 - \frac{\partial \varepsilon_i}{\partial \psi_s}\right) \qquad \text{Eqn. 1c}$$

$$= C_Q\left(1 - \frac{\partial \varepsilon_i}{\partial \psi_s}\right) \qquad \text{Eqn. 1d}$$

The first part of Eqn. 1c, $C_Q$, is the quantum capacitance, which is a measure of the density of states at the Fermi level. $C_S$ is degraded from $C_Q$ by a factor that is proportional to how much $\varepsilon_i$ (the difference of the $i^{th}$ subband to $E_c$) changes. Ideally, at high inversion conditions $\varepsilon_i$ should be constant, meaning that the quantized levels and $E_c$ shift by same amount and the subband levels can easily get in the potential well that forms at the Si/SiO$_2$ interface. This directly translates on the wavefunction been able to come closer to the interface as the surface is inverted more and more. However, $\varepsilon_i$ can float up as charge accumulates in the device, giving rise to the differential term in Eqn. 1d, and the wavefunction stays away from the interface. As shown earlier on in Fig. 2b, this shift is almost 0.5nm. Other than the wavefunction shift, $C_Q$ being small is the second degrading factor of $C_S$ as indicated in Eqn. 1d. Figure 4b shows the $C_Q$ of the three nanowires as a function of $V_G$, calculated as the density of states at the Fermi level. Clearly, for all wires the maximum value is below 3nF/nm, not even an order of magnitude above $C_{OX}$=0.48nF/nm. The fact, that the position of the charge distribution degrades $C_S$ from $C_Q$ by almost four times, ($C_S = 0.8 nF/m$), indicates its large significance on the device's capacitance. (Similar deviations of the semiconductor capacitance from the quantum capacitance have also been observed in thin body devices [35]).

*Variations in $C_Q$ between different wire orientations:* As shown in Fig. 4b, in all wire cases, $C_Q$ is not constant, but undergoes large transitions as the Fermi level is pushed inside the subbands at large gate biases. This is expected, since $C_Q$ is a measure of the density of states at the Fermi level, and the differences in the dispersion cause differences in $C_Q$. Comparing $C_Q$ for different wire orientations, the [111] wire has the largest $C_Q$ for most of the bias range because of the higher mass (m*=0.47$m_0$) and higher degeneracy of its valleys (D=6). The $C_Q$ drop at high biases in the [111] case is associated with the decreasing 1D density of states away from the band edges, and due to the fact that its bands flatten out at Γ and do not extent as parabolic bands in k-space as shown in Fig. 3d. On the other hand, the [100] and [110] wires initially have lower $C_Q$, because of their lower density of states (lighter masses and lower degeneracies). At high biases, the upper valleys of the [100] and [110] wires start to get populated, which allows a continuous increase in $C_Q$ for these wires. More specifically, since the charge in all cases is almost the same at a given bias, the same number of states in each wire need to be occupied. The Fermi level in the [110] wire with lower mass and smaller valley degeneracies reaches the upper valleys faster (at a lower gate bias) than the [100] wire in order to occupy the same number of states. Once this happens, the $C_Q$ of the [110] wire surpasses the $C_Q$ of the [100] wire (around $V_G$=0.4V).

*Variations in $C_Q$ do not cause variations in $C_G$:* The differences in $C_Q$, between wires in different orientations, however, are not large enough to result in differences in the total capacitances. As seen earlier, $C_Q$ is only partially responsible for the total capacitance degradation. The small differences in $C_Q$ are smeared out in $C_G$ by the oxide capacitance, and the charge shift from the interface, that is very similar for all the above wires. (This observation can of course be different in the case of high-k dielectric oxides, in which the importance of $C_Q$ can be more pronounced).

*Velocity controls the transport differences in different orientated wires:* As explained above, the charge is almost the same in all three nanowires. Since in the ballistic limit the ON-current performance is given by the product of "charge *times* velocity", if the charge is the same, any performance differences will result from differences in the carrier velocities. Figure 4c shows the injection velocities of the wires vs. gate bias ($V_G$). The [110] wire has the largest velocities whereas the [111] wire has the lowest velocities. In all cases, the injection velocities are not constant, but increase as the lattice is filled with charge because faster high energy carrier states are being populated. This increase in velocities, calculated form the initial value at low gate biases to the final value at high gate biases can reach up to 17% in the [110] wires and even up to 27% in the [100] and 24% in the [111] wire orientation cases. When comparing the velocities of the different wires, however, the masses of the valleys determine the velocities of the carriers. (In 1D, under the parabolic band approximation, the velocity is proportional to $v \sim 1/\sqrt{m^*}$). As a result, the [110] wire with m*=0.16m$_0$ has the highest velocity, followed by the [100] wire with mass m*=0.27m$_0$, and finally by the [111] wire of mass m*=0.47m$_0$. The larger density of states of the [111] wire and its larger degeneracy do not allow the Fermi level to be pushed far into the conduction band. Therefore, only the lower energy and slower carries are used, and the velocity in this case is low. In the [110] wire case, the degeneracy is 2, and the subband density of states low, therefore the Fermi level will be pushed far into the conduction band, and faster carries will be utilized as shown in Fig. 4c.

*Velocity differences affect the I-V differences:* The velocity difference directly reflects on the $I_{DS}$ as shown in Fig. 4d in which the drive current capabilities of the wires are compared at the same $I_{OFF}$. The [110] and [100] wires perform better than the [111] wire in terms of ON-current capabilities. The current in the [110] wire stands ~5% higher than the [100] wire and ~20% higher than the [111] wire because of its lower mass. This result must be qualified since the bandstructure of the wires is a very sensitive function of their quantization. The results presented here are for these specific 3nm wire examples. In cases where important dispersion parameters such as the relative placement of the valleys in energy, masses and degeneracies, are altered, different conclusions might be drawn, especially for the relative performance of the [100] and [110] wires which is not that large. In the next section, an analysis is performed on how exactly these parameters (valleys splittings that lift degeneracies, and masses) are affected by quantization.

*(c) Quantization influence on valley splitting and mass variation*

Quantization strongly affects both factors that control the performance, the degeneracies and masses. In this section of the paper, the effect of quantization on these parameters is examined. Degeneracies are controlled mainly by the orientation, but can be lifted due to valley splitting [12-14] under strong quantization (both electrostatic and structural).

*Weak valley splitting in [100] and [111] quantized wires:* Figure 5a-b shows the $E(k)$ of a 2nm wire in the [100] and [111] orientations. A slight valley splitting of the degenerate valleys under quantization is observed. In the case of the [100] the splitting is 10meV and in the case of the [111] wire, 24meV. These values are less than the room temperature $k_BT$=26meV and are not expected to have a significant effect in the transport properties of the nanowires at room temperature.

*Strong valley splitting in [110] quantized wires:* In the case of [110] nanowires, valley splitting is significantly larger. As shown in Fig. 5c in the $E(k)$ of a 2nm [110] wire, Γ and off-Γ valleys experience valley splitting of their degeneracies by 76meV and 14meV respectively. Figure 5d shows how this effect varies with the spatial confinement in the [110] wire. Although large nanowires (>5nm) are not affected, the valleys splitting can reach up to 200meV for the Γ valleys of narrow wires with sizes as narrow as 1.5nm. The valley splitting of the off-Γ valleys, on the other hand, is not affected as much. Only a few tenths of meVs of splitting are observed in this case. (It is noted here that the splitting in the other wire orientations is smaller than the [110] wires of similar quantization sizes even for wires below 2nm [14]).

*Generally, masses increase with increase in quantization:* The effective mass is the second important transport performance dispersion property that is affected by quantization of the nanowire cross section. The injection velocity and quantum capacitance strongly depend on the masses. Both the quantization and the transport masses of nanowires under arbitrary wire orientations are certain combinations of the longitudinal ($m_l$=0.89$m_0$) and the transverse effective masses ($m_t$=0.19$m_0$) of the Si ellipsoids. Figure 6a shows the three pairs of ellipsoids that form the conduction band minima in Si, each characterized by the *x, y* and *z* directional masses. The masses of the valleys that appear in the nanowire dispersion are automatically included in tight-binding. What will be shown is that under quantization, the exact values of these masses are changed from their bulk values. In most cases, quantization results in an increase in the effective mass. Figure 6b shows the variation in the lowest valley transport masses as the dimension of the wire cross section reduces. At large wire cross sections, the mass of the the [100] valley that is located at Γ, approaches the bulk transverse mass $m_t$=0.19$m_0$. The bulk mass of the [111] wire is larger since it is a combination of $m_t$=0.19$m_0$ and $m_l$=0.89$m_0$ (the bulk value is 0.43$m_0$) [18, 36]. The mass in the [100] case almost doubles as the dimension of the wire's side decreases from 7.1nm to 1.5nm (88% increase). (The 3nm wire has $m^*$=0.27$m_0$ as mentioned earlier). The corresponding increase in the [111] wire's mass is 17%, with the 3nm wire having $m^*$=0.47$m_0$. The off-Γ valley masses (upper valleys) of both [100] and [110] wires also increase as the dimension reduces as shown in Fig. 6c. In the [100] off-Γ valley case, a slight mass increase of 9% between the 7.1nm and the 1.5nm is extracted from the bandstructure calculations. The off-Γ valley mass increase in the [110] case is 11%.

*[110] wire Γ valley masses decrease with increase in quantization:* In contrast to the rest of the valleys, the Γ valley mass of the [110] oriented wires decreases with increase in quantization. As shown in Fig. 6b the mass decreases by 32% as the side of the wire reduces from 7.1nm to 1.5nm. As mentioned earlier, the mass of a 3nm [110] wire is $m^*$=0.16$m_0$, which gives an enhanced injection velocities and transport characteristics of [110] wires over the rest of the wires. Anisotropy and non-parabolicity in the Si conduction band Brillouin zone cause this unintuitive behavior as explained in the next section.

*(d) Understanding the nanowire mass variation as a function of quantization.*

*Semi-analytical construction of the wire's dispersion:* This distinctly different observation in the masses of wires is a result of the non-parabolicity and anisotropy of the Si bandstructure. Under any physical quantization, the subband levels will follow the "particle in a box" quantization, as shown in Fig. 7a. The smaller the physical domain, the larger the corresponding quantized *k*, and the higher the energy levels of the subbands. To estimate the quantization levels of the Si conduction band ellipsoid quantized along the longitudinal direction, the energy contour in the *x-y* plane near the band minima is plotted in Fig. 7b. ("Cut" through the ellipsoid along its longitudinal axis). Similarly to Fig. 7a, quantization of *Lx* of 2nm, 3nm, and 5nm will shift the energy levels to the vertical lines shown in the figure. The energy levels at these lines will be the relevant subbands in an ultra-thin-body (UTB) quantization – with one quantized dimension. Figure 7c now, shows the energy contour taken at the 3nm line, perpendicular to the contour of Fig. 7b in the *y-z* plane. An extra quantization in the *z*-direction (the second quantized dimension, as in the wire case) will leave only one allowed *k*-space variable, the transport direction one. This forms the 1D dispersion of the wire. The relevant 1D bands are the ones located at the horizontal lines of Fig. 7c. Lines for *Lz*=2nm, 3nm and 5nm are shown. The solid line indicates a relative subband for an UTB device with *Lx*=3nm and $Lz = \infty$ ($k_z$=0, only one quantization dimension).

*Mass and band edge extraction from the semi-analytical construction:* The 1D subbands of Fig. 7c are plotted in Fig. 7d for the cases of *Lz*=2nm, 3nm and 5nm. (The *x*-direction quantization is *Lx*=3nm in all cases). The mass of these bands is the transport mass (*y*-direction) that the wire has in the [010] orientation (equivalent to the [100] wire orientation described in the previous sections). Smaller cross sections raise the subband energy, and increase the masses. Through this process, both, the transport masses and the placement of the subband edges in energy can be deduced. From the subband edges the quantization masses can be extracted. The more non-parabolic the bulk bandstructure is at higher energies in the direction of quantization, the slower the subbands rise in energy with quantization compared to the parabolic band case. This results in larger quantization masses. The more non-parabolic the bulk bandstructure is in the transport direction, the larger the transport masses will be. All these effects appear in thin body channel devices (UTB of Fig. 6d), however, they are significantly more enhanced in the case of nanowires because of the extra quantization of one more physical dimension [19, 30].

*Different orientations, different anisotropies:* The transport masses of wires in other orientations can be explained similarly. Evident in the bandstructure is the anisotropy which results in different behavior in the quantization of the [100] to quantization of the [110] axes. In [110] oriented wires, the [100] and [0-11] directions are quantized. The [100] quantization is the same step as the one in Fig. 7b. Quantizing the [0-11] direction, will result in extracting 1D bands by lines that cross Fig. 7d at 45°, (in $L_{yz}$, instead of horizontal). Figure 7e shows the first subband of the dispersions of structures with $Lx$=3nm and $Lyz$=2nm, 3nm, 5nm and $\infty$, similar to Fig. 7c. Evident in this case is the non-parabolicity of the dispersion, as it is also evident in Fig. 5c. For comparison purposes, Fig. 7g shows the positive $k_{yz}$ branch of the dispersion, with all the bands shifted to the origin. Clearly, as the structure is quantized in the [0-11] direction, the curvature of the dispersion increases, corresponding to a lowering of the transport mass of the wires. In contrast to the [001] quantization case of Fig. 7c, here the anisotropy in the bandstructure results in a reduction of the transport masses with increase in quantization, in agreement with the calculation for the actual nanowire mass shown in Fig. 6b. The magnitude of the mass variation is however smaller in the [0-11] quantization direction compared to the [001] direction. (Similar anisotropic results have been also obtained using empirical non-local pseudopotential and ab-initio GW calculations [21]).

*Limitations of the semi-analytical construction:* This construction method can provide a rough guidance as to what the dispersion of a nanowire will look like. The method, however, does not include any of the interactions between the bands/valleys (which are enhanced when the material is physically confined in a nanowire), and lacks any band coupling information. Effects such as valley splitting, that are a consequence of band coupling, cannot be captured. The extracted mass values, as well as their variation trends under quantization, are however quite accurate. In the case of nanowire electronic transport for nanowires larger than 3nm, where the mass is an important transport parameter, a first order estimation of the nanowires' performance can be drawn by using this analytical mass extraction.

# 4. Conclusions

Transport properties of nanowires in different transport orientations ([100], [110] and [111]) were examined using a 10 orbital $sp^3d^5s^*$ atomistic TB model self consistently coupled to a 2D Poisson solver. A semiclassical ballistic model was used to calculate the current-voltage characteristics of the nanowires. The dispersions of the nanowires undergo changes under gate bias, which at some cases can cause large lift of degeneracies and small subband shifts. Although these changes under self-consistency do not alter the velocity and density of states of the wires, they are associated with the spatial distribution of charge that together with the small 1D density of states can degrade the nanowire's capacitance by 30%. The quantum capacitance of the different oriented 3nm wires that were investigated is a strong function of gate bias, but of similar magnitude in all wires. Almost the same is also the total gate capacitance of all nanowire devices in different orientations investigated as well as the inversion charge. Due to their lighter mass, 3nm [110] oriented wires have the maximum injection velocities, whereas [111] oriented wires the lowest injection velocities due to their higher masses. The injection velocity reflects directly on the current capabilities of the wires, where the [110] and [100] oriented wires indicate the best performance in terms of ON-current capabilities compared to the [111] wires which are the worst. The masses of the wires are a sensitive function of the wire dimensions (below 7nm), and strongly influence the output performance of nanowire devices. This is an effect that resides in the non-parabolicity and anisotropy of the Si Brillouin zone that is particularly important in strongly quantized devices. Valley splitting is another effect strongly dependent on quantization. [110] nanowires of dimensions below 3nm are extremely sensitive to this. Finally, the authors would like to mention that the simulator used in this study will be released as an enhanced version of the Bandstructure Lab on nanoHUB.org [37]. This simulation engine allows any user to duplicate the simulation results presented here. Over 800 users have utilized the Bandstructure Lab in the past 12 months.

## 5. Acknowledgements

This work was funded by the Semiconductor Research Corporation (SRC) and MARCO MSD Focus Center on Materials, Structures and Devices. The computational resources for this work were provided through nanoHUB.org by the Network for Computational Nanotechnology (NCN). The authors would like to acknowledge Prof. Mark Schilfgaarde of Arizona State University for ab-initio GW calculations and Dr. Tony Low of Purdue University for pseudopotential calculations for benchmarking of bandstructure results, and Prof. Timothy Boykin of University of Alabama at Huntsville for tight-binding discussions.

# References

[1]  ITRS Public Home Page. http://www.itrs.net/reports.html

[2] N. Singh et. al., "Ultra-narrow silicon nanowire gate-all-around CMOS devices: Impact of diameter, channel-orientation and low temperature on device performance," *Int. Elec. Dev. Meeting*, 2006.

[3] K. H. Cho, "Observation of single electron tunneling and ballistic transport in twin silicon nanowire MOSFETs (TSNWFETs) fabricated by top-down CMOS process," *Int. Elec. Dev. Meeting*, 2006.

[4] J. Xiang, W. Lu, Y. Hu, Y. Wu, H. Yan, and Charles M. Lieber, "Ge/Si nanowire heterostructures as high-performance field-effect transistors," *Nature*, vol. 441, no. 25, 2006.

[5] T. B. Boykin, G. Klimeck, and F. Oyafuso, "Valence band effective-mass expressions in the sp3d5s* empirical tight-binding model applied to a Si and Ge parametrization," *Phys. Rev. B*, vol. 69, pp. 115201-115210, 2004.

[6] G. Klimeck, F. Oyafuso, T. B. Boykin, R. C. Bowen, and P. von Allmen, *Computer Modeling in Engineering and Science (CMES)*, vol. 3, no. 5, pp. 601-642, 2002.

[7] G. Klimeck, S. Ahmed, H. Bae, N. Kharche, S. Clark, B. Haley, S. Lee, M. Naumov, H. Ryu, F. Saied, M. Prada, M. Korkusinski, and T. B. Boykin, "Atomistic simulation of realistically sized nanodevices using NEMO 3-D—Part I: Models and benchmarks," *IEEE Trans. Elecron Devices*, vol. 54, no. 9, pp. 2079-2089, 2007.

[8] J. C. Slater and G. F. Koster, "Simplified LCAO method for the periodic potential problem," Phys. Rev., vol 94, no. 6, 1954.

[9] M.S. Lundstrom, and J. Guo, "Nanoscale transistors: Device physics, modeling and simulation," *Springer*, 2006.

[10] A. Rahman, J. Guo, S. Datta, and M. Lundstrom, "Theory of ballistic nanotransistors," *IEEE Trans. Electron Devices*, vol. 50, no. 9, pp. 1853-1864, 2003.

[11] M. Luisier, A. Schenk, W. Fichtner, "Full-Band Atomistic Study of Source-To-Drain Tunneling in Si Nanowire Transistors," Proc. of SISPAD, 978-3-211-72860-4, pp. 221-224, 2007.

[12] T. B. Boykin, G. Klimeck, M. Friesen, S. N. Coppersmith, P. von Allen, F. Oyafuso, and S. Lee, "Valley splitting in strained silicon quantum wells," *Appl. Phys. Lett.*, vol. 84, no. 1, pp.115, 2004.

[13] T. B. Boykin, G. Klimeck, M. Friesen, S. N. Coppersmith, P. von Allen, F. Oyafuso, and S. Lee, "Valley splitting in low-density quantum confined heterostructures studied using tight-binding models," *Phys. Rev. B*, vol. 70, pp.165325, 2004.

[14] A. Rahman, G. Klimeck, M. Lundstrom, T. B. Boykin, and N. Vagidov, "Atomistic approach for nanoscale devices at the scaling limit and beyond – Valley splitting in Si ," J. J. Appl. Phys., vol. 44, no. 4B, pp. 2187-2190, 2005.

[15] J. Wang, A. Rahman, A. Ghosh, G. Klimeck, and M. Lundstrom, "On the validity of the parabolic effective-mass approximation for the I–V calculation of silicon nanowire transistors," *IEEE Trans. Electron Devices*, vol. 52, no. 7, pp. 1589-1595, 2005.

[16] M. Luisier, A. Shenk, W. Fichtner, and G. Klimeck, "Atomistic simulations of nanowires in the sp3d5s* tight-binding formalism: From boundary conditions to strain calculations," Phys. Rev. B, vol. 74, no. 20, p. 205323, Nov. 2006.

[17] E. Gnani, S. Reggiani, A. Gnudi, P. Parruccini, R. Colle, M. Rudan, and G. Baccarani, "Band-structure effects in ultrascaled silicon nanowires," IEEE Trans. Electron Devices, vol. 54, no. 9, 2007.

[18] M. Bescond, N. Cavassilas, and M. Lannoo, "Effective-mass approach for n-type semiconductor nanowire MOSFETs arbitrarily oriented," *Nanotechnology*, vol. 18, pp. 255201, 2007.

[19] Y. Liu, N. Neophytou, T. Low, G. Klimeck, and M. Lundstrom, "A tight-binding study of the ballistic injection velocity for ultra-thin-body SOI MOSFETs," *IEEE Trans. Electron Devices,* to appear.

[20] N. Neophytou, A. Paul, M. Lundstrom and G. Klimeck, "Simulations of nanowire transistors: Atomistic vs. effective mass models," J. Comput. Electr., 2007, to appear.

[21] Personal communication with Prof. Mark Schilfgaarde (Arizona State University) for the GW calculations and with Dr. Tony Low (Purdue University) for the pseudopotential calculations. A comparison between the methods will be published in the future.

[22] R. C. Bowen, G. Klimeck, R. Lake, W. R. Frensley and T. Moise, "Quantitative Resonant Tunneling Diode Simulation," *J. of Appl. Phys.* 81, 3207 (1997)

[23] G. Klimeck, T. B. Boykin, R. C. Bowen, R. Lake, D. Blanks, T. Moise, Y. C. Kao, and W. R. Frensley, "Quantitative Simulation of Strained InP-Based Resonant Tunneling Diodes," Proceedings of the 1997 55th IEEE Device Research Conference Digest, IEEE, NJ, p. 92 (1997)

[24] R. C. Bowen, C. Fernando, G. Klimeck, A. Chatterjee, D. Blanks, R. Lake, J. Hu, J. Davis, M. Kularni, S. Hattangady, and I.C. Chen, "Physical Oxide Extraction and

Versification using Quantum Mechanical Simulation," Proceedings of IEDM 1997, IEEE, 869 (1997).

[25] D.D.D. Ma et.al., "Small-diameter silicon nanowire surfaces," *Science,* 299, 1874, 2003.

[26] J. Wang, PhD thesis, Purdue University, 2005.

[27] X. Zhao, C. M. Wei, L. Yang, and M. Y. Chou, "Quantum confinement and electronic properties of silicon nanowires," *Phys. Rev. Lett.*, vol. 92, no. 23, pp. 236805, 2004.

[28] N. Kharche, M. Prada T. B. Boykin and G. Klimeck, "Valley splitting in strained silicon quantum wells modeled with 2° miscuts, step disorder, and alloy disorder," Appl. Phys. Lett. 90, 092109, 2007.

[29] R. Rahman, C. J. Wellard, F. R. Bradbury, M. Prada, J. H. Cole, G. Klimeck, and L. C. L. Hollenberg, "High precision quantum control of single donor spins in silicon," Phys. Rev. Lett. 99, 036403, 2007.

[30] G. Liang, J. Xiang, N. Kharche, G. Klimeck, C. M. Lieber, and M. Lundstrom, "Performance Analysis of a Ge/Si Core/Shell Nanowire Field Effect Transistor", Nano Lett., vol. 7, 642-646 (2007),

[31] S. Lee, F. Oyafuso, P. Von, Allmen, and G. Klimeck, "Boundary conditions for the electronic structure of finite-extent embetted semiconductor nanostructures," *Phys. Rev. B*, vol. 69, pp. 045316-045323, 2004.

[32] We have conducted extensive tests to validate the 2D Poisson solution compared to the actual 3D solution. A maximum deviation of 2% on band edges between the two approaches was found, and ignorable deviation in the I-V characteristics. However, the

2D method reduces the Poisson computational time by almost 5X compared to the 3D solution.

[33] The length of the unit cell in [111] is $L_{111} = \sqrt{3}a_0$. Therefore, the Brillouin zone of a 1D nanowire in [111] extends from $-\frac{\pi}{\sqrt{3}a_0}$ to $\frac{\pi}{\sqrt{3}a_0}$. The $\Delta$ valleys in bulk Si are located at $k_x = 0.815 \cdot 2\pi / a_0$. Under quantization in (111) surface they project on the [111] axis at $k_{[111]} = 0.815 \cdot 2\pi /\left(\sqrt{3}a_0\right) = 1.63\pi /\left(\sqrt{3}a_0\right)$. The valley projection point falls in the 2[nd] Brillouin zone. It is then folded in the first Brillouin zone as $k_{[111]} = 1.63\pi /\left(\sqrt{3}a_0\right) - 2\pi /\left(\sqrt{3}a_0\right) = 0.37\pi /\left(\sqrt{3}a_0\right)$. After normalization to the length of the Brillouin zone, the valleys appear at $k_{[111]} = 0.37$.

[34] N. Neophytou, A. Paul, M. Lundstrom and G. Klimeck, "Self-consistent simulations of nanowire transistors using atomistic basis sets," Proc. of SISPAD, 978-3-211-72860-4, pp. 217-220, 2007.

[35] H. S. Pal, K. D. Cantley, Y. Liu & M. S. Lundstrom, "Semiconductor capacitance and injection velocity considerations for bulk and SOI MOSFETs with alternate channel materials," *IEEE Trans. Electron Devices,* 2008, to appear.

[36] A. Rahman, M. S. Lundstrom, and A. W. Ghosh, "Generalized effective-mass approach for n-type metal-oxide-semiconductor field-effect transistors on arbitrarily orientated wafers," *J. Appl. Phys.*, vol. 97, pp. 053702-053713, 2005.

[37] [nanoHub] Bandstructure lab on nanoHUB.org
(https://www.nanohub.org/tools/bandstrlab/)

# Figure Captions

Figure 1:

(a) Simulation procedure schematic. Using an atomistic $sp^3d^5s^*$-SO tight-binding model, the bandstructure of the nanowire under consideration is calculated. A semiclassical ballistic model is then used to calculate the charge distribution in the wire from the source and drain Fermi levels. The charge is used in a 2D Poisson for the electrostatic solution of the potential in the cross section of the wire. The whole process is done self consistently. (b) The lattice in the wire transport orientations (surfaces) used – [100], [110] and [111].

Figure 2:

Device features for a 3nm [100] rectangular wire. (a-b) The 2D cross section showing the charge distribution under low and high gate biases, respectively. Even under high bias, the charge distribution is located almost half a nanometer away from the oxide. This causes degradation in the total capacitance of the wire. The dots indicate the underlying atomic positions. (c-d) $E(k)$ plots for the cases (a-b). The bandstructure features change under self consistency. $E_{fs}$ is the source Fermi level. (Zero energy indicates the conduction subband edge.)

Figure 3:

(a-b) The bandstructure of a 3nm [110] oriented nanowire under low bias ($V_G$=0V) (a) and high bias ($V_G$=0.8V) (b), and $V_D$=0.5V. Under high biases the degeneracies of the Γ valley are lifted from 2 to 1. (c-d) The bandstructure of a 3nm [111] oriented nanowire under low bias ($V_G$=0V) (c) and high bias ($V_G$=0.8V) (d) and $V_D$=0.5V. (e) The charge in the wire as a function of the difference of the conduction band edge from the Fermi level for two cases: (1) The Fermi level "scans" the equilibrium bandstructure and the charge is extracted, and (2) the charge is extracted from the self-consistent calculations with potential variations in the lattice taken into consideration. (f) The injection velocity for the same case as (e). The changes in the dispersion themselves do not reflect much on the charge distribution or the injection velocities. The differences between the two models result form the spatial information of the wavefunction that corresponds to the bandstructure changes.

Figure 4:

Performance comparison of the 3nm square wires in the [100], [110] and [111] directions at the same OFF-current ($I_{OFF}$) (a) The gate capacitance $C_G$ vs. gate bias ($V_G$). The capacitance is similar for all wires, and degraded from the oxide capacitance by an amount that corresponds to an increase in the effective oxide thickness of 0.54nm. (b) The quantum capacitance $C_Q$ vs. $V_G$ of the three devices, which is a measure of the density of states at the Fermi level. (c) Comparison between the injection velocities of the nanowires vs. $V_G$. In all cases, the velocity is not constant, but increases as the gate bias increases. The increase is calculated by the difference between the value at high $V_G$ and the value at low $V_G$. (d) The $I_{DS}$ vs. $V_G$ for the three wires at the same $I_{OFF}$. The velocity difference directly reflects on the current differences.

Figure 5:

The effect of valley splitting in small nanowires. (a-c) The $E(k)$ of a 2nm wire in the [100], [111] and [110] orientations respectively. (d) The effect of valley splitting in the [110] wire as the dimensions decrease. The Γ valleys are severely affected at cross sections below 3nm, whereas the off-Γ valleys are not affected as much.

Figure 6:

(a) The three equivalent pairs of ellipsoids in the conduction band of Si are described by the longitudinal and transverse masses. A combination of these masses results in the quantization and transport masses of nanowires under arbitrary orientations. (b) The transport masses oriented in [100], [110] and [111] vs. wire dimension as calculated from TB. At large wire cross sections, the [100] and [110] that are located at Γ, approach the bulk $m_t$=0.19$m_0$. The mass of the [111] wire is larger since it is a combination of $m_t$ and $m_l$=0.89$m_0$. As the wire dimensions shrink, the mass of the [110] wire reduces, whereas the masses of the other two wires increase. (c) The off-Γ valley masses for the cases of the [110] and [100] wires. Both increase as the dimensions decrease. (The expected bulk mass values for every orientation are denoted). The percentage change denoted is the change in the effective masses between the 1.5nm mass value (mostly scaled wire) and the 7.1nm wire.

Figure 7:

(a) The energy levels of a quantized structure using the "particle in a box" picture. Under quantization, the subband edges and masses can be deduced from the materials' bulk dispersion with a numerical $E(k)$ diagram. (b) Energy contour at the middle of one of the Si Brillouin zone ellipsoids calculated using the full 3D k-space information of the Si Brillouin zone. A "cut" through the Si ellipsoid along its longitudinal axis is shown. Under quantization in $Lx$=2nm, 3nm and 5nm, the relevant subband energies are indicated by the vertical $k_x$=constant lines. (c) A "cut" through the Si ellipsoid perpendicular to its longitudinal axis at the $k_x$ line corresponding to the 3nm quantization line of (b). The non-parabolicity and anisotropy is evident in this figure. The horizontal lines indicate the relevant energy regions under another quantization in the y-direction for $Ly$=2nm, 3nm and 5nm quantized structures. The solid line labeled UTB is the relevant band for an ultra-thin-body (UTB) device of $Lx$=3nm thickness in [001] with $Lz = \infty$. This is only quantized in the x-direction. (d) The dispersions of the vertical lines in (c). The masses and the band edge of the dispersions will be the ones that appear in a quantized wire. (e) The 2D plot is the same as in (c). The 45° lines correspond to a quantization in the [0-11] for $Lyz$=2nm, 3nm and 5nm. The solid line labeled UTB is the relevant band for an ultra-thin-body (UTB) device of $Lx$=3nm thickness in [001] with $Lyz = \infty$. (f) The dispersions of the 45° lines in (e). The non-parabolicity is evident in this orientation. (g) Zoom of the right (positive momentum) branch of (f) with all dispersions shifted to the origin for comparison. As the structure is quantized in [0-11], the mass becomes lighter. The anisotropy in the Brillouin zone is directly reflected on the masses in the different wire orientations (as in Fig. 6b).

Figure 1

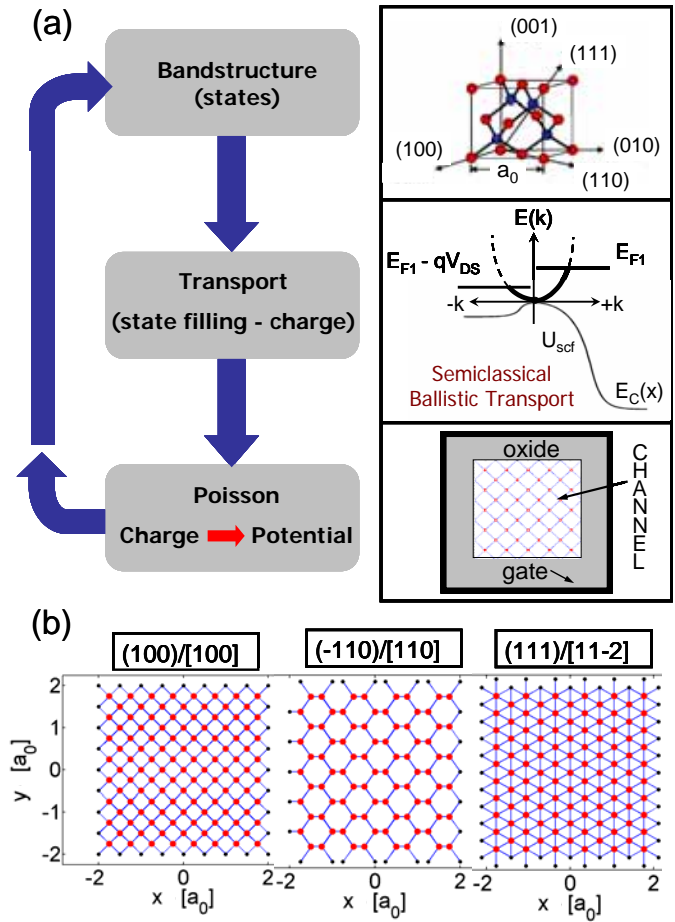

Figure 2

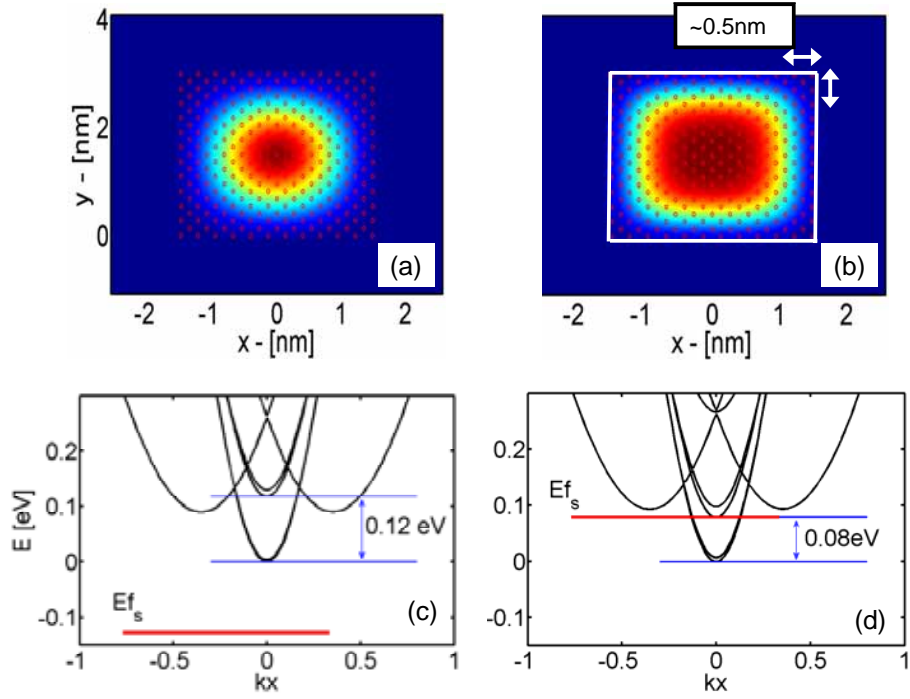

Figure 3

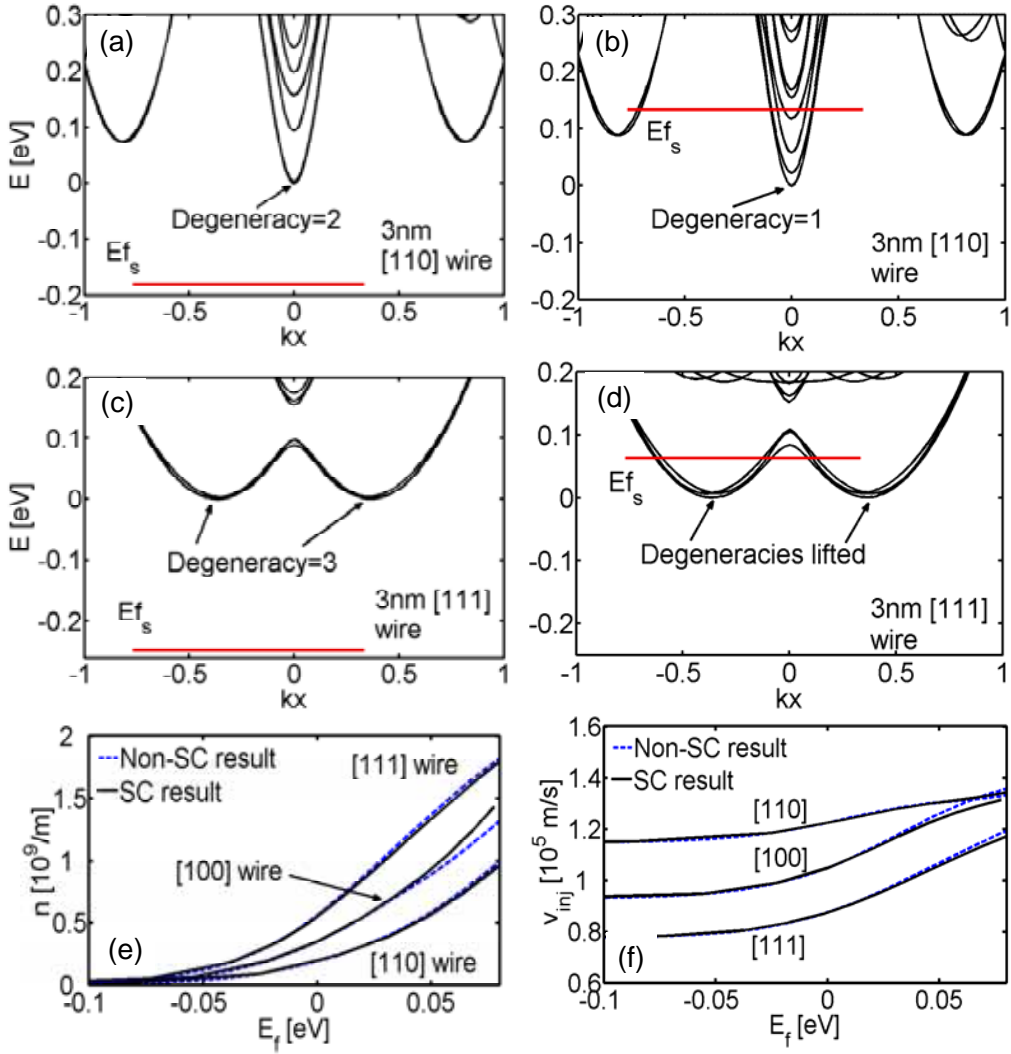

Figure 4

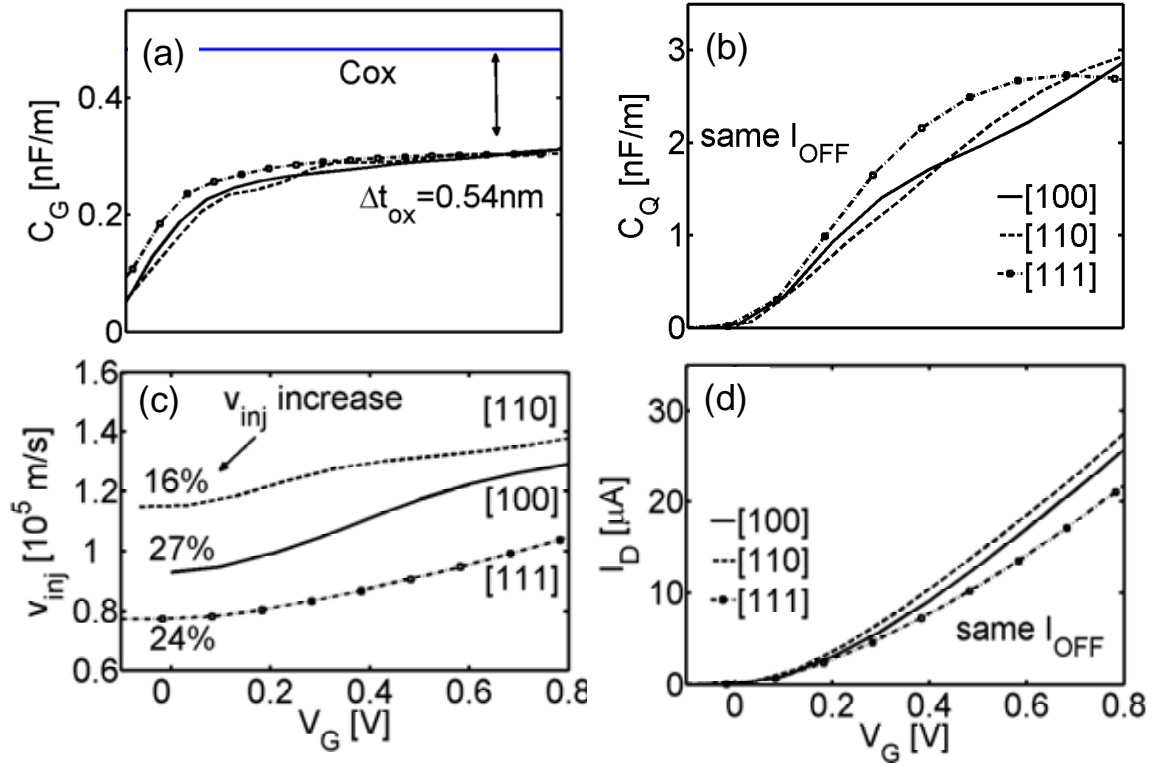

Figure 5

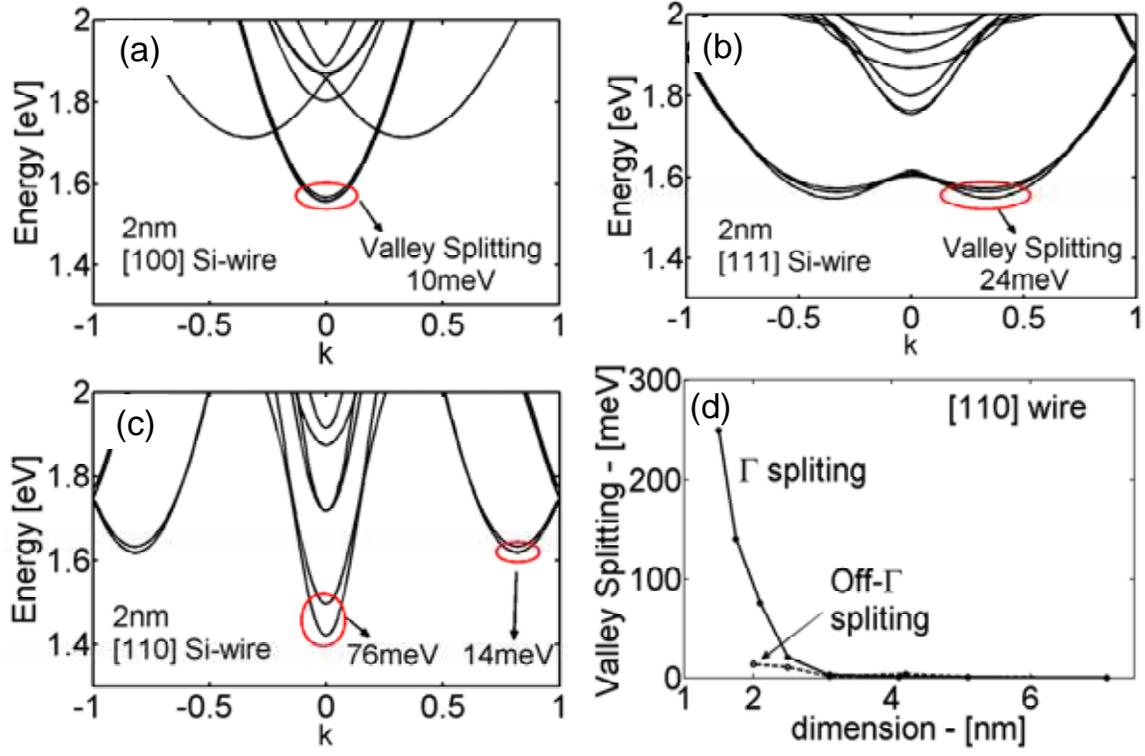

Figure 6

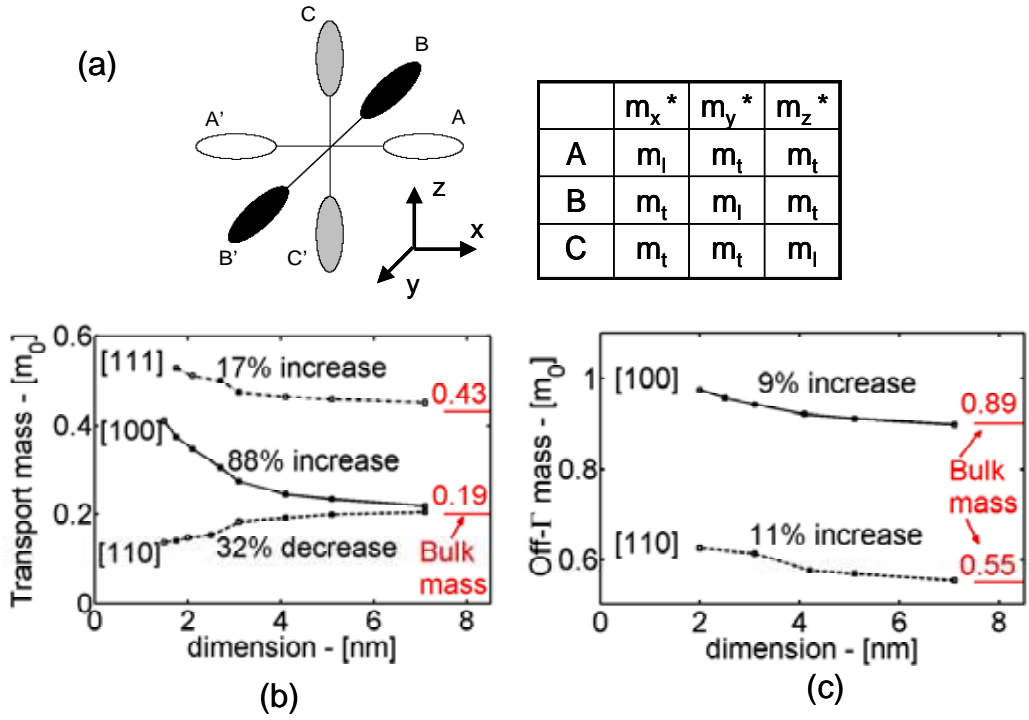

(a)

(b)

(c)

Figure 7

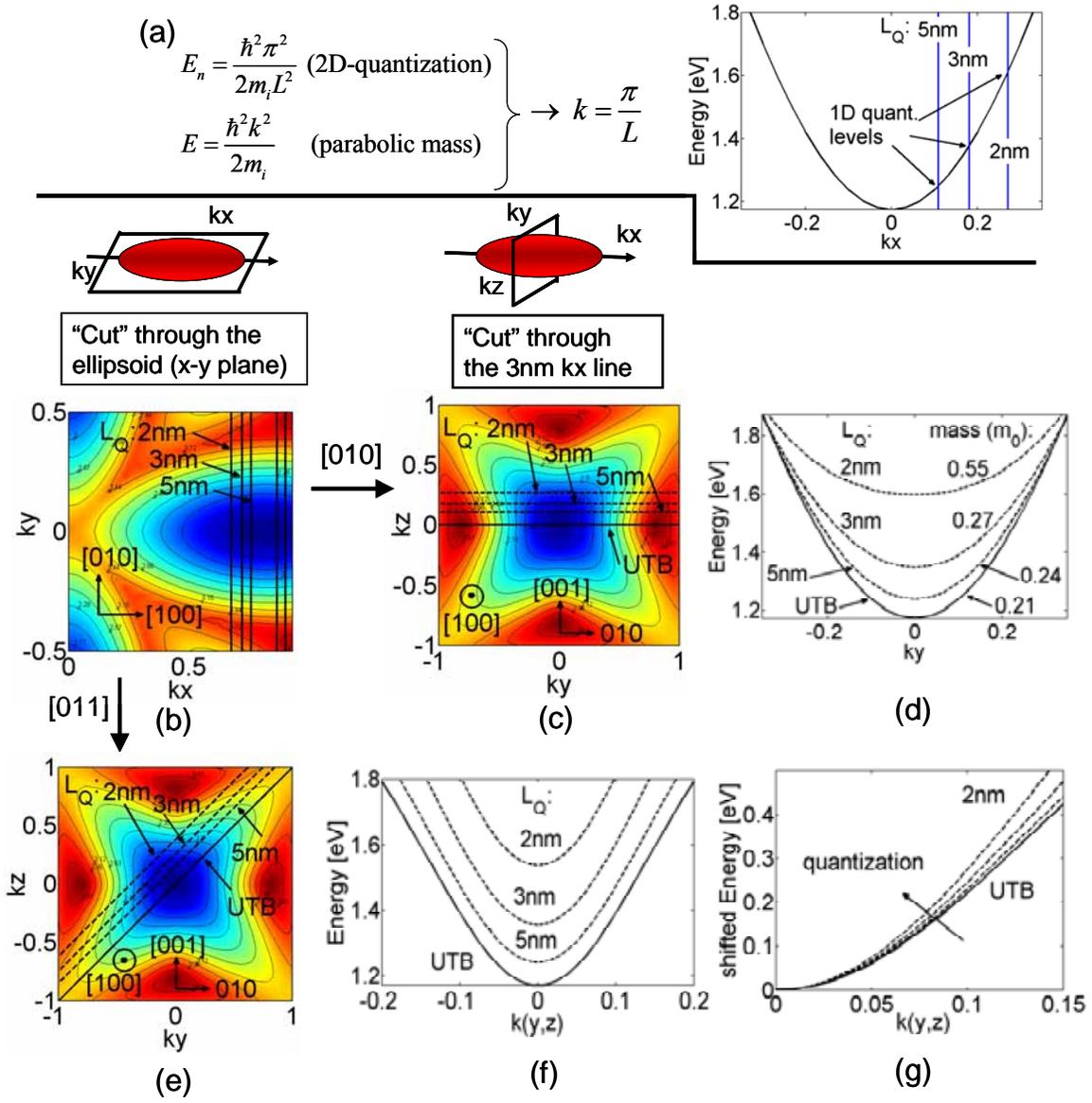